\documentclass[a4paper]{jpconf}

\usepackage{graphicx}
\usepackage{amsmath}
\usepackage{color}
\usepackage{siunitx}

\begin{document}
\title{Neutrino oscillation tomography of the Earth with KM3NeT-ORCA}

\author{Simon Bourret\textsuperscript{1}, Jo\~ao A B Coelho\textsuperscript{1} and V\'eronique Van Elewyck\textsuperscript{1,2}\\
on behalf of the KM3NeT collaboration}

\address{\textsuperscript{1} APC, Universit\'e Paris Diderot, CNRS/IN2P3, Sorbonne Paris Cit\'e, F-75205 Paris, France}
\address{\textsuperscript{2} Institut Universitaire de France, 75005 Paris, France}
\ead{bourret@apc.in2p3.fr}

\begin{abstract}
KM3NeT-ORCA is a water-Cherenkov neutrino detector designed for studying the oscillations of atmospheric neutrinos, with the primary objective of measuring the neutrino mass ordering.
Atmospheric neutrinos crossing the Earth undergo matter effects, modifying the pattern of their flavour oscillations. The study of the angular and energy distribution of neutrino events in ORCA can therefore provide tomographic information on the Earth's interior with an independent technique, complementary to the standard geophysics methods.
Preliminary estimations based on a full Monte Carlo simulation of the detector response show that after ten years of operation the electron density can be measured with a precision of 3-5\% in the mantle and 7-10\% in the outer core -- depending on the mass ordering.
\end{abstract}
\section{Introduction and motivations}

ORCA (\textit{Oscillation Research with Cosmics in the Abyss}) is the low-energy branch of KM3NeT, the next-generation neutrino Cherenkov detector currently being built in the Mediterranean Sea with the aim of measuring the neutrino mass ordering and searching for high-energy cosmic neutrino sources~\cite{Adrian-Martinez2016}.
The ORCA detector will instrument 5.7\,Mton of seawater in a dense configuration of 115 vertical strings with an horizontal spacing of 20\,m, anchored on the seabed off the shore of Toulon (France). Each string supports 18 digital optical modules with a vertical spacing of 9\,m.  With this configuration, ORCA will focus on the study of atmospheric neutrino oscillations in the energy range $\sim\!1\mbox{-}100\text{\,GeV}$.

Due to coherent forward scattering on electrons, the flavour oscillations of atmospheric neutrinos propagating through the Earth matter are modified with respect to vacuum oscillations: this is known as the MSW effect~\cite{Mikheev1985,Wolfenstein1978}. An accurate measurement of this effect, based on the angular, energy and flavour distribution of neutrino interactions in ORCA, can provide tomographic information on the electron density in the Earth's interior~\cite{Rott2015,Winter2016}. This method is complementary to usual geophysical methods such as inversion of seismic data and geodetic measurements, which are used to infer the radial mass density profile. The ratio of electron to mass density scales with the average proton to nucleon ratio (hereafter denoted $Z/A$), which depends on the chemical composition of the medium. Hence, combining both measurements may allow us to constrain compositional models of the inner Earth -- e.g. in the outer core~\cite{Rott2015}. In this work we use a reference geophysical model of the radial mass density profile in the Earth and we estimate the sensitivity of ORCA to the $Z/A$ ratio in the mantle and outer core.

\section{Methodology}

Neutrino event topologies in ORCA can be separated into two broad classes, for which dedicated reconstruction algorithms have been developed. Track-like events result from charged-current (CC) interactions of $\nu_{\mu} / \overline{\nu_{\mu}}$ and $\nu_{\tau}/ \overline{\nu_{\tau}}$ which produce an outgoing $\mu / \overline{\mu}$. Cascade-like events correspond to all other interaction channels: charged-current interactions of $\nu_{e} / \overline{\nu_{e}}$ and $\nu_{\tau}/ \overline{\nu_{\tau}}$ and neutral-current (NC) interactions, producing only hadronic and electromagnetic showers.
Event classification, together with rejection of down-going atmospheric muons, are based on a Random Decision Forest algorithm. We refer to~\cite{Adrian-Martinez2016} for details regarding simulations, detector resolutions and classification performance.

\begin{figure}[h]

\begin{minipage}{0.56\linewidth}

The trajectory of an atmospheric neutrino crossing the Earth is modelled by a \textit{baseline} consisting of a finite number of steps of constant electron density (Fig.\,\ref{fig:PREM}). A set of baselines corresponding to the angular binning is derived from a radial model of the Earth with 42 density layers, where mass density values are fixed and follow the PREM model~\cite{Dziewonski1981}. Additionnally, three chemical layers are defined, where the composition, and hence the $Z/A$ factor, are assumed to be uniform:
\begin{itemize}
\setlength\itemsep{0em}
\item[(a)] solid iron inner core: $R = 0-1221\,\text{km}$%, Z/A = 0.466$
\item[(b)] liquid iron outer core: $R = 1221-3480\,\text{km}$%, Z/A = 0.466$
\item[(c)] silicate mantle (and crust): $R = 3480-6368\,\text{km}$ %\\this layer comprises the whole silicate Earth (mantle and crust)
\end{itemize}

\end{minipage}
\hspace{0.02\linewidth}
\centering
\begin{minipage}{0.40\linewidth}
\includegraphics[width=\linewidth]{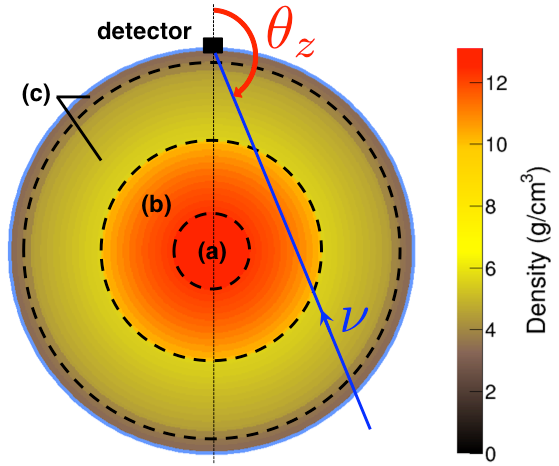}
\caption{\label{fig:PREM}\footnotesize Density layers of the PREM.}
\end{minipage}

\end{figure}

The composition measurement is extracted from the  energy $E$ and zenith angle $\theta_{z}$  (see Fig.\,\ref{fig:PREM}) distributions of track-like and cascade-like events, in bins of reconstructed $\log E$ and $\cos\theta_z$. To compute the rate of interacting events in the detector, we use atmospheric flux tables from the HKKM2014 simulations~\cite{Honda2015}, and neutrino-nucleon cross-sections weighted for water molecules from the GENIE Monte Carlo generator~\cite{Andreopoulos2009}.
\vspace{-1ex}
\vspace{-1ex}

Using custom software, the evolution equation for neutrino propagation in matter is solved numerically for each step in order to obtain the probabilities of flavour transition.
The mass ordering is assumed to be known at the time of the tomography measurement, and oscillation parameters are fixed to their global (ordering-dependent) best-fit values as of May 2016~\cite{Gonzalez-Garcia2014}.

The detector response $(E_{\text{true}}, \cos\theta_{\text{true}})\rightarrow(E_{\text{reco}}, \cos\theta_{\text{reco}})$ and flavour identification are applied using binned 4-dimensional response matrices built from Monte Carlo simulated events, accounting for detection and reconstruction efficiencies, event classification performance and errors on reconstructed variables (including correlations). We consider 16 distinct channels: $\{ \text{CC }\nu_{e} / \overline{\nu_{e}}, \text{ CC }\nu_{\mu} / \overline{\nu_{\mu}}, \text{ CC }\nu_{\tau}/ \overline{\nu_{\tau}}, \text{ NC } \nu / \overline{\nu} \} \times \{ \text{track-like, cascade-like}\}$.
The contamination of signal channels by atmospheric muons is of the order of a few percent and is not included due to limited Monte Carlo statistics.

\section{Preliminary results and prospects}
At this stage, the reported sensitivities account for statistical uncertainty only -- the influence of systematic effects is currently under study. We use a binned log-likelihood ratio test statistic combining both signal channels:
$$
\Delta \chi^2 = \sum\limits_{\substack{\text{Tracks,} \\ \text{Cascades}}} \,
\sum\limits_{\substack{\text{bins\,}\log E \\ \text{\,bins\,} \cos\theta_{z} } }
 2 \Big[ n_{\text{exp}} - n_{\text{obs}} + n_{\text{obs}} \cdot \ln \Big( \frac{n_{\text{obs}}}{n_{\text{exp}}} \Big) \Big]
$$
where $n_{\text{exp}}$ denotes the expected number of events under a certain Earth model hypothesis (model experiment), and $n_{\text{obs}}$ is the observed number of events under an assumption on the true Earth parameters (pseudo-data). In this simplified analysis the sensitivity can be evaluated from the average experiment or \textit{Asimov dataset}, i.e. no pseudo-experiments are drawn.
\begin{figure}[h]
\centering
\begin{minipage}{0.48\linewidth}
\includegraphics[width=1.12\linewidth]{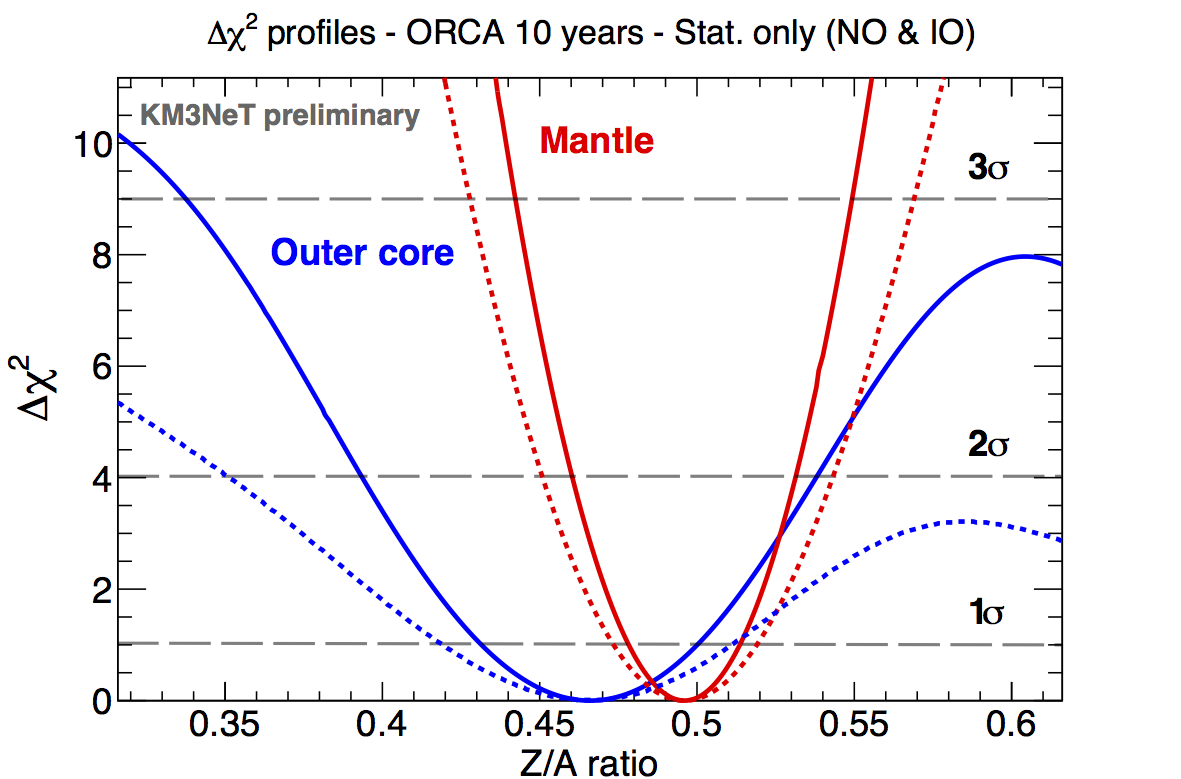}
\caption{\footnotesize{\label{fig:DeltaChi2}$\Delta\chi^{2}$ profiles for mantle and outer core. Solid lines: normal ordering assumed. Dotted lines: inverted ordering assumed.}}
\end{minipage}\hspace{0.04\linewidth}%
\begin{minipage}{0.48\linewidth}
\includegraphics[width=1.12\linewidth]{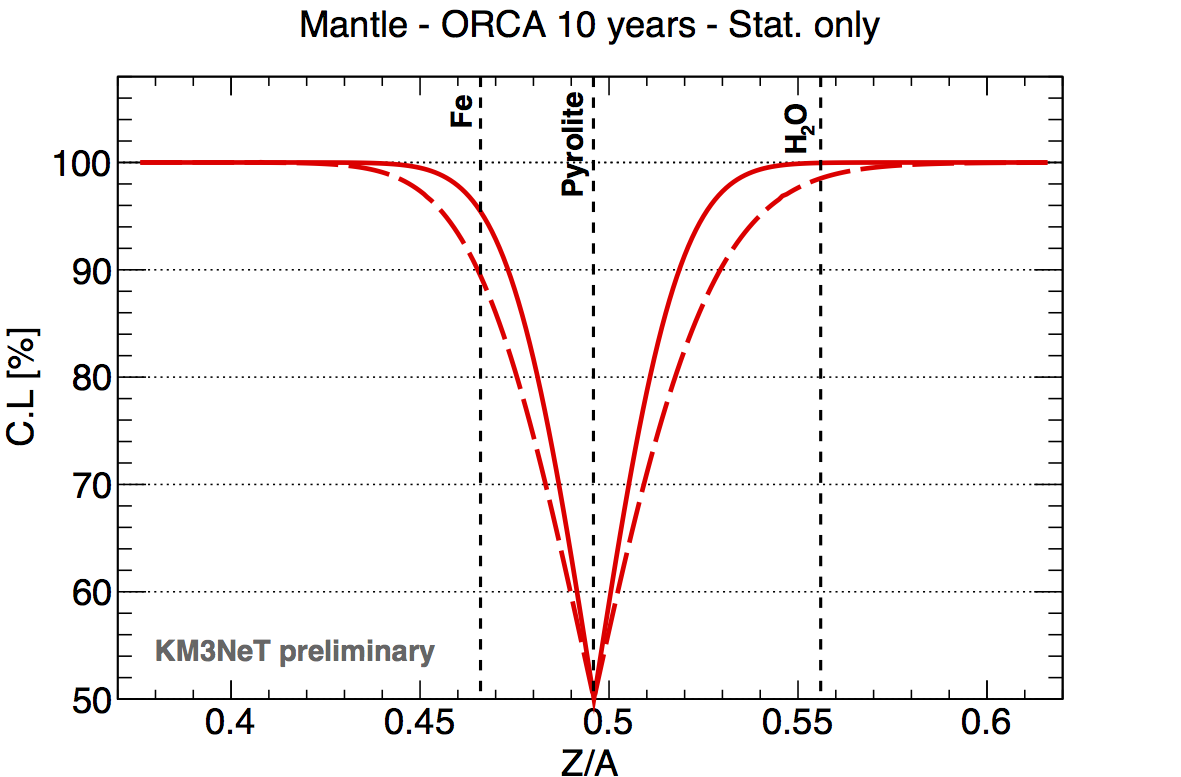}
\caption{\footnotesize{\label{fig:MantleVPlot}
Confidence level for rejecting the pyrolitic mantle hypothesis (normal ordering).
Solid line: combined channels. Dashed line: track-like only.
}}
\end{minipage}
\end{figure}

Fig.\,\ref{fig:DeltaChi2} shows that ORCA will be more sensitive to the electron density in the mantle than in the core. Although the oscillations are enhanced for core-crossing neutrinos, the resonance energy ($\sim 2\text{-}4\,\text{GeV}$) is closer to the detection threshold and fast oscillations cannot be well resolved in this region due to limited angular and energy resolutions. High-density values for the outer core cannot be excluded with strong significance, due to self-similarities of the oscillation probabilities as the electron density increases~\cite{Winter2016}. The composition of the mantle is usually approximated by a theoretical rock model called \textit{pyrolite} ($Z/A = 0.496$). We report the estimated confidence level for rejecting this hypothesis as a function of the alternative true $Z/A$ in Fig.\,\ref{fig:MantleVPlot}.

A confidence interval for the measurement of the $Z/A$ factor for both layers can be derived from the 1-dimensional $\Delta \chi^2$ profiles -- considering the two measurements independently. After 10 years of operation, we estimate that the proposed ORCA detector will be able to measure the electron density in the Earth mantle to an accuracy of $\pm3.6\%$ (resp.\,$\pm4.6\%$) at $1\sigma$ confidence level, assuming normal (resp. inverted) mass ordering. In the outer core, this accuracy is reduced to $\pm7.4\%$ (resp.\,$\pm10.0\%$). As shown in Fig.\,\ref{fig:MantleVPlot}, a significant improvement is realised by combining track-like and cascade-like channels.
Further studies will include improved reconstructions of the interaction inelasticity, which can help to distinguish neutrinos from antineutrinos on a statistical basis, as well as the impact of systematic effects and multi-layer simultaneous measurement.

{\footnotesize
\subsection*{Acknowledgements}
The authors gratefully acknowledge financial support from the IdEx and LabEx UnivEarthS programs at Sorbonne Paris Cit\'e (ANR-11-IDEX-0005-02, ANR-10-LABX-0023).
}

\section*{References}

\bibliographystyle{iopart-num}
\bibliography{biblio_bibtex}

\end{document}